# Which Adaptation Logic? An Objective and Subjective Performance Evaluation of HTTP-based Adaptive Media Streaming Systems

Christian Timmerer
Alpen-Adria-Universität Klagenfurt
Universitätsstraße 65-67
9020 Klagenfurt
+43-463-2700-3621
christian.timmerer@itec.aau.at

Matteo Maiero
Alpen-Adria-Universität Klagenfurt
Universitätsstraße 65-67
9020 Klagenfurt
+43-463-2700-3600
mmaiero@edu.uni-klu.ac.at

Benjamin Rainer
Alpen-Adria-Universität Klagenfurt
Universitätsstraße 65-67
9020 Klagenfurt
+43-463-2700-3627
benjamin.rainer@itec.aau.at

## ABSTRACT
Multimedia content delivery over the Internet is predominantly using the Hypertext Transfer Protocol (HTTP) as its primary protocol and multiple proprietary solutions exits. The MPEG standard Dynamic Adaptive Streaming over HTTP (DASH) provides an interoperable solution and in recent years various adaptation logics/algorithms have been proposed. However, to the best of our knowledge, there is no comprehensive evaluation of the various logics/algorithms. Therefore, this paper provides a comprehensive evaluation of *ten different adaptation logics/algorithms,* which have been proposed in the past years. The evaluation is done both objectively and subjectively. The former is using a predefined bandwidth trajectory within a controlled environment and the latter is done in a real-world environment adopting crowdsourcing. The results shall provide insights about which strategy can be adopted in actual deployment scenarios. Additionally, the evaluation methodology described in this paper can be used to evaluate any other/new adaptation logic and to compare it directly with the results reported here.

## Keywords
Dynamic Adaptive Streaming over HTTP, Performance Evaluation, Crowdsourcing, Subjective Quality Assessment, Quality of Experience, QoE, DASH, MPEG

## 1. INTRODUCTION
Multimedia content is omnipresent in our daily life and we consume it (among others) with different devices and in various contexts ranging from wired to wireless connections on large, high-resolution screens and small mobile devices. In many cases the content is no longer stored on the actual device but streamed from servers (within a cloud) over the open, unmanaged Internet. The Hypertext Transfer Protocol (HTTP) is nowadays considered as the primary protocol for the delivery of multimedia content over the Internet and various approaches have been proposed, starting with download-and-play, progressive download, and, recently, adaptive HTTP streaming. For the latter, various proprietary solutions are deployed from notable companies but with MPEG's Dynamic Adaptive Streaming over HTTP (DASH) a standardized solution is in place which offers interoperability among different vendors [1].

The basic design principle of DASH (and its proprietary predecessors) is that multimedia content is provided in various versions (e.g., different bitrates, resolutions, qualities, etc.), which are referred to as *representations*. These versions are divided into equally sized and time-aligned *segments* which can be located using HTTP uniform resource locators (HTTP-URLs) and independently downloaded using a conventional HTTP access client. A DASH client receives a manifest describing the relationship among representations and other metadata, which is an XML document and referred to as *Media Presentation Description (MPD)*. The DASH client is now free to instruct the HTTP access client to download segments from any representation contained within the MPD in order of its appearance and to concatenate the media segments at the client in order to reconstruct a continuous media presentation. By doing so, the DASH client may switch to different representations during the streaming session according to context conditions observed during download of individual segments (e.g., changes in the available bandwidth or even changes in the device). The component, which is typically responsible for deciding on these representation switches, is generally referred to as adaptation logic and not defined within the standard but deliberately left open for competition.

Since the ratification of the MPEG-DASH standard in 2011 and its official publication by ISO in 2012, many research papers have been published addressing various aspects of adaptive HTTP streaming including proposals for the adaptation logic and its evaluation under various conditions or the comparison of different approaches. For example, Akhshabi et al. provides a first evaluation of rate-adaptation algorithms in adaptive streaming over HTTP [2]. Therefore, they investigate two commercial players (i.e., Microsoft Smooth Streaming and Netflix) and one open source player (OSMF), which are evaluated under different conditions. Müller et al. [15] perform a similar evaluation but in vehicular environments using real-world bandwidth traces captured while driving on a highway and accessing multimedia content over HTTP. In particular, they compare Microsoft Smooth Steaming, Adobe HTTP Dynamic Streaming, and Apple HTTP Live Streaming to their own DASH-based implementation which uses a simple throughput-based adaptation logic. A similar comparison is conducted by Riiser et al. [19] within 3G networks, again using bandwidth traces collected in real-world trials, which uses OSMF, Apple HTTP Live Streaming, Microsoft Smooth Steaming, and Netview's Media Client. All the above evaluations focus on objective metrics like throughput, start-up delay, or number of stalls and a limited number of DASH-like clients equipped with different adaptation logics. For commercial deployments, these adaptation logics are typically not accessible and only black box testing is possible.

In practice, however, to the best of our knowledge and at the time of writing this paper, no research paper provides a comprehensive evaluation of various adaptation logics/algorithms including a methodology that makes it easy to compare others (including those not yet developed) with results reported in the literature. Therefore, this paper evaluates *ten different adaptation logics/algorithms,* which have been proposed in the past years. In particular, the evaluation is conducted both objectively and subjectively. The objective evaluation is conducted within a controlled environment based on established metrics known in the literature. Additionally, we define the inefficiency and instability of the adaptation logic as a derived metric. The subjective evaluation is conducted within a real-world environment using crowdsourcing.

The remainder of this paper is organized as follows. Section 2 provides an overview of the adaptation logics used in this paper. The evaluation methodology for both objective and subjective evaluation is described in Section 3. The actual results are presented in Section 4 – for the objective evaluation – and Section 5 – for the subjective evaluation –, respectively. Finally, Section 6 concludes the paper.

## 2. OVERVIEW OF ADAPTATION LOGICS

This section introduces the adaptation logics used in this paper in order to understand their characteristics, behaviors, advantages, and drawbacks. The following ten adaptation logics have been evaluated in this paper and are briefly introduced in the following: DASH-JS [3], FESTIVE [4], Instant [5], Liu et al. [6], Miller et al. [7], OSMF [8], PANDA [9], QDASH [10], Thang et al. [11], and Tian-Liu [12].

We acknowledge the existing of other adaptation logics but an exhaustive comparison is probably not feasible. However, this paper and its methodology can be used as a basis to compare any other adaptation logic (existing ones and those yet to be developed) with the results obtained in this evaluation due to the open access approach adopted in this work. That is, test conditions and all material required to reproduce these tests are publicly available. Finally, the terms adaptation logic and adaptation algorithm are used interchangeably in this paper.

### 2.1 DASH-JS

*DASH-JS* is one of the first implementations adopting the W3C Media Source Extensions (MSE) which allow a seamless integration within the Web environment [3]. It is based on a simple bandwidth estimation as shown in Equation ( 1 ).

$$b_n = \frac{w_1 b_{n-1} + w_2 b_m}{w_1 + w_2} \qquad ( 1 )$$

where $b_{n-1}$ is the throughput calculated at the $n-1^{th}$ segment, $b_m$ denotes the throughput measured during the download of the $n-1^{th}$ segment, while $w_1$ and $w_2$ are weighting factors that adjust the influence of the recently measured segment download (i.e., $w_1$=0.7 and $w_2$=1.3 according to [3]). Thus, the bandwidth estimated for the next segment is calculated taking 1.3 times the bitrate observed for the last segment downloaded and 0.7 times the estimated throughput that was calculated during the previous call. The initialization is based on the bandwidth measured when downloading the MPD.

### 2.2 FESTIVE

*FESTIVE* is a Fair, Efficient, Stable, adaptIVE algorithm which is one of the first algorithms taking into account interactions across multiple adaptive streaming players that compete for bandwidth [4]. Therefore, FESTIVE introduces different components to reach its goal, i.e., a harmonic bandwidth estimator, a stateful and delayed bitrate update, and a randomized scheduler.

The segment requests are randomized over the timeline, allowing for a fair share of the available bandwidth. The algorithm switches only to the next higher/lower representation (i.e., bitrate) available according to the MPD, with the proposed bitrate for the next segment that is calculated relating the current bitrate with the throughput of the bandwidth estimation that FESTIVE adopts as a smoothed value computed over the last n segments with n=20 as in [4]. The bitrate for the next segment is calculated based on a given cost function, which provides a balance among efficiency, fairness, and stability.

### 2.3 Instant

The *Instant* algorithm simply takes the bandwidth measured during the download of the last segment which is mapped to the available representation (i.e., bitrate) according to the MPD [5]. In particular, the representation which bitrate is lower than the measured bandwidth is used for the next segment request. The initialization is done in the same way as for DASH-JS by using the measured bandwidth while downloading the MPD.

### 2.4 Liu et al.

The algorithm proposed by Liu et al. [6] – in the following simply referred to as *Liu* – is one of the first adaptation logics in this domain. The algorithm is based on a smoothed HTTP/TCP throughput measurement method that relates the segment fetch time with the media playback time contained in that segment. It is in some way similar to TCP's Additive Increase Multiplicative Decrease (AIMD) where switching up to a higher representation is additive (i.e., to the next higher bitrate) and switching down is multiplicative (i.e., to the actual observed bandwidth which possibly skips some representations). When the two thresholds are not met, the algorithm keeps the selected rate.

### 2.5 Miller et al.

The goal of this algorithm – henceforth referred to as *Miller* – is to adapt the bitrate requested depending on the buffer level and trying to maximize the average bitrate while minimizing the number of quality switches [7]. It uses the available throughput calculated for the previous segment and the available buffer level as an input, providing the representation (i.e., bitrate) for the next segment and the minimum buffer level when to start the download of the segment as an output. The algorithm can be divided in two phases: an initial fast start-up phase that increases the quality of the downloaded segments in a more aggressive manner and a stable condition where the algorithm prefers to keep a high buffer level. Finally, it keeps multiple thresholds ($0 < b_{min} < b_{low} < b_{max}$) with the objective to keep the buffer level in an optimal range defined as $b_{opt}= 0.5(b_{low}+b_{max})$.

### 2.6 OSMF

The Open Source Media Framework (*OSMF*) is provided by Adobe [8] and comes with a very basic adaptation logic, based on a factor calculated as the ratio between the media segment duration time and the time needed to download that segment. It provides instant reaction to bandwidth changes and allows for skipping intermediate representations, i.e., it is possible to switch immediately to the highest or lowest quality representation whereas others typically adopt a step-wise switching approach.

Table 1. MPEG-DASH Representations for Test Sequence with Representation ID, Resolution [pixels], and Bitrate [kbps].

| Rep.id | Res. [px] | Bitrate [kbps] | Rep.id | Res. [px] | Bitrate [kbps] |
|---|---|---|---|---|---|
| 1 | 192×108 | 100 | 9 | 1920×1080 | 1300 |
| 2 | 192×108 | 150 | 10 | 1920×1080 | 1600 |
| 3 | 320×180 | 200 | 11 | 1920×1080 | 1900 |
| 4 | 480×270 | 350 | 12 | 1920×1080 | 2300 |
| 5 | 960×540 | 500 | 13 | 1920×1080 | 2800 |
| 6 | 960×540 | 700 | 14 | 1920×1080 | 3400 |
| 7 | 960×540 | 900 | 15 | 1920×1080 | 4500 |
| 8 | 1280×720 | 1100 | | | |

## 2.7 PANDA

*PANDA* stands for Probe-AND-Adapt, which means that the actions of probing and rate adapting are the basic principles of this algorithm [9]. It basically probes the network by incrementing the request rate preparing to back off when congestion is experienced. This constant network probing has the advantage that competing clients will observe the correct status of the network in a few steps and hopefully share the available resources in a fair way.

## 2.8 QDASH

*QDASH* takes into account the Quality of Experience (QoE) based on the assumption that users typically do not notice quality improvements while they heavily criticize quality degradation [10]. Therefore, QDASH adopts a step-wise adaptation to lower quality representations in case of a bandwidth drop in order to mitigate this effect. The original version of this algorithm adopts a proxy service for bandwidth estimation, which is replaced here – for simplicity and without impacting its performance – with an instantaneous evaluation of the available bandwidth performed during segment download.

## 2.9 Thang et al.

The algorithm proposed by Thang et al. aims for a smooth playback during short-term bandwidth fluctuations but reacts quickly in case of larger bandwidth drops [11]. Therefore, the *Thang* algorithm uses a sliding average of the observed media throughput, which dynamically adapts to changing bandwidth conditions. Interestingly, the first segment to be requested always belongs to the lowest representation, which is in contrast to others that use the MPD download as an educated guess, e.g., [3][5].

## 2.10 Tian-Liu

Finally, the algorithm proposed by Tian and Liu [12] – in this paper simply referred to as *Tian-Liu* – uses the buffer to mitigate bandwidth fluctuations and enable smooth playback. Additionally, if the available bandwidth drops below a certain threshold, the adaptation logic starts to behave like Instant in order to leverage the difference between throughput observed and bitrate selected, in order to avoid buffer underruns/stalls and restore a sufficient video buffer level.

## 3. METHODOLOGY

### 3.1 Introduction

This section describes the methodology for evaluating the adaptation logics introduced in Section 2.

The test sequence is based on the DASH dataset [13] where we adopt the Big Buck Bunny sequence that we encoded with ffmpeg and segmented with GPAC's MP4Box [14] in order to get the representations as shown in Table 1. The configuration is inspired by [15] and provides a good mix of resolutions and bitrates for both fixed and mobile network environments. In fact, we provide two versions, one with a segment length of 2s and the other with 10s that are the most common segment sizes currently adopted by actual deployments (i.e., Apple HLS uses 10s whereas others like Microsoft and Adobe use 2s).

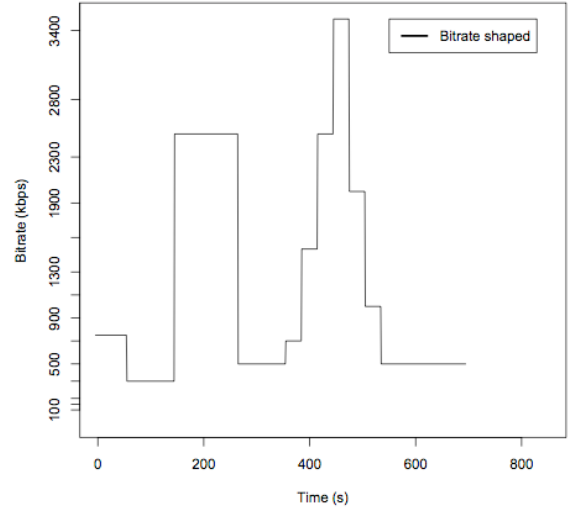

Figure 1. Bandwidth Trajectory for Objective Evaluation within a Controlled Environment.

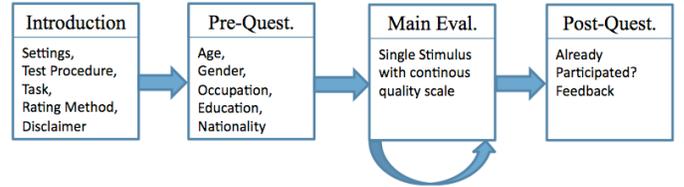

Figure 2. Subjective Evaluation Methodology.

### 3.2 Objective Evaluation Setup

For the objective evaluation we adopt the setup according to [15] where the bandwidth and delay between a server and client are shaped using a shell script, that invokes the Unix program TC with netEM and a token bucket filter. In particular, the delay was set to 80ms and the bandwidth follows a predefined trajectory as shown in Figure 1. The delay corresponds to what can be observed within long-distance fixed line connections or reasonable mobile networks and, thus, is representative for a broad range of application scenarios. The bandwidth trajectory contains both abrupt and step-wise changes in the available bandwidth to properly test all the adaptation logics under different conditions.

The actual shell script is attached to the paper as supplemental material to enable reproducible research.

The goal of this evaluation setup is to provide objective metrics which are collected at the client to be analyzed during the evaluation. These metrics include the observed bitrate, selected quality representation, buffer level, start-up delay, stalls (re-buffering due to underruns), and derived metrics as detailed in Section 4.

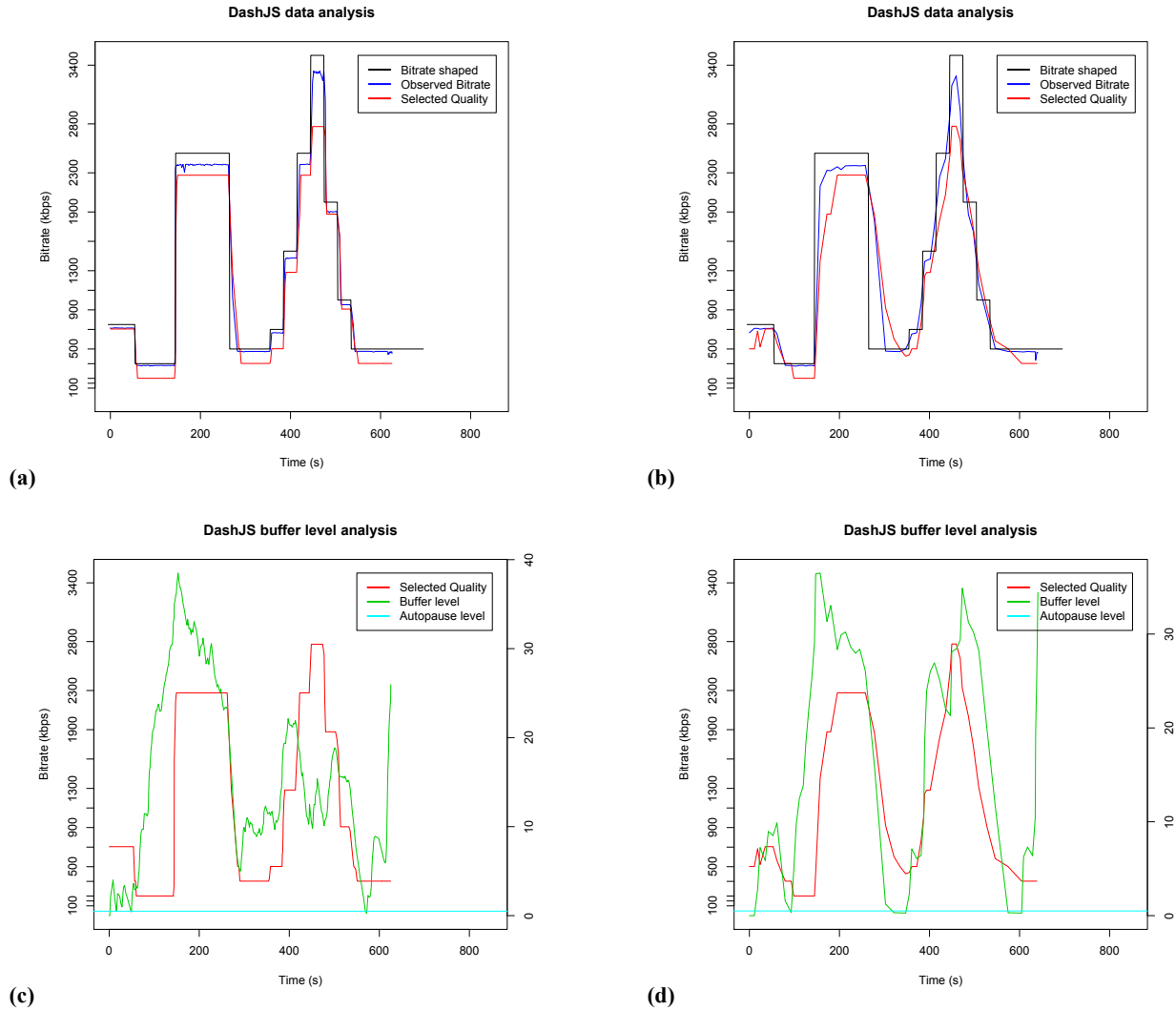

**Figure 3. Bandwidth Adaptation for DASH-JS with 2s and 10s segment length: (a) adaptation logic 2s, (b) adaptation logic 10s, (c) buffer level 2s, (d) buffer level 10s. The buffer file state is provided in seconds on the right side.**

## 3.3 Subjective Evaluation Setup

For the subjective evaluation we adopt a crowdsourcing approach according to [16] that uses the Microworker platform to run such campaigns and to recruit participants, which are actually referred to as microworkers. The content server is located in Europe and, thus, we limit participants to Europe in order to reduce network effects due to proxies, caches, or content distribution networks (CDNs) that we cannot control as identified in [16].

At the end of the subjective evaluation, each microworker needs to hand in a proof that she/he has successfully participated which is implemented using a unique identification number. We set the compensation to US$ 0.4, which is the minimum compensation for this type of campaign at the time of writing this paper (we noticed an increase in compensation required by the platform over time).

The stimulus is the same as for the objective evaluation but we added another sequence – an excerpt from Tears of Steel, also available at [13] – in order to mitigate any bias that may be introduced when using only one type of content. The content configuration is the same as shown in Table 1 but we used only one segment size of 2s.

Figure 2 depicts the subjective evaluation methodology comprising an *introduction*, a *pre-questionnaire*, the *main evaluation*, and a *post-questionnaire*. The introduction explains the structure of the task and how to assess the actual QoE asking the microworker to provide a honest response. The pre-questionnaire collects demographic data like country of residence that we use later to filter participants. The main evaluation comprises a Web site presenting the stimulus (both sequences) with a gray background as recommended in [17]. The content is actually streamed over the open Internet to which the microworker is connected using a JavaScript-based DASH client with one of the adaptation logics as described in Section 2. The selection of the adaptation logic is uniformly distributed (p=1/10) among the participants and the size of DASH client is fixed to a resolution of 1280×720 pixels. After the stimulus presentation, participants rate the QoE using a slider with a continuous scale from 0 to 100. The slider is initially set to 50 (middle position) and the time for rating the QoE is limited to eight seconds [17]. The stimulus – both sequences – is presented in random order to the participants. Finally, the post-questionnaire gathers any feedback from the participants using a free text field.

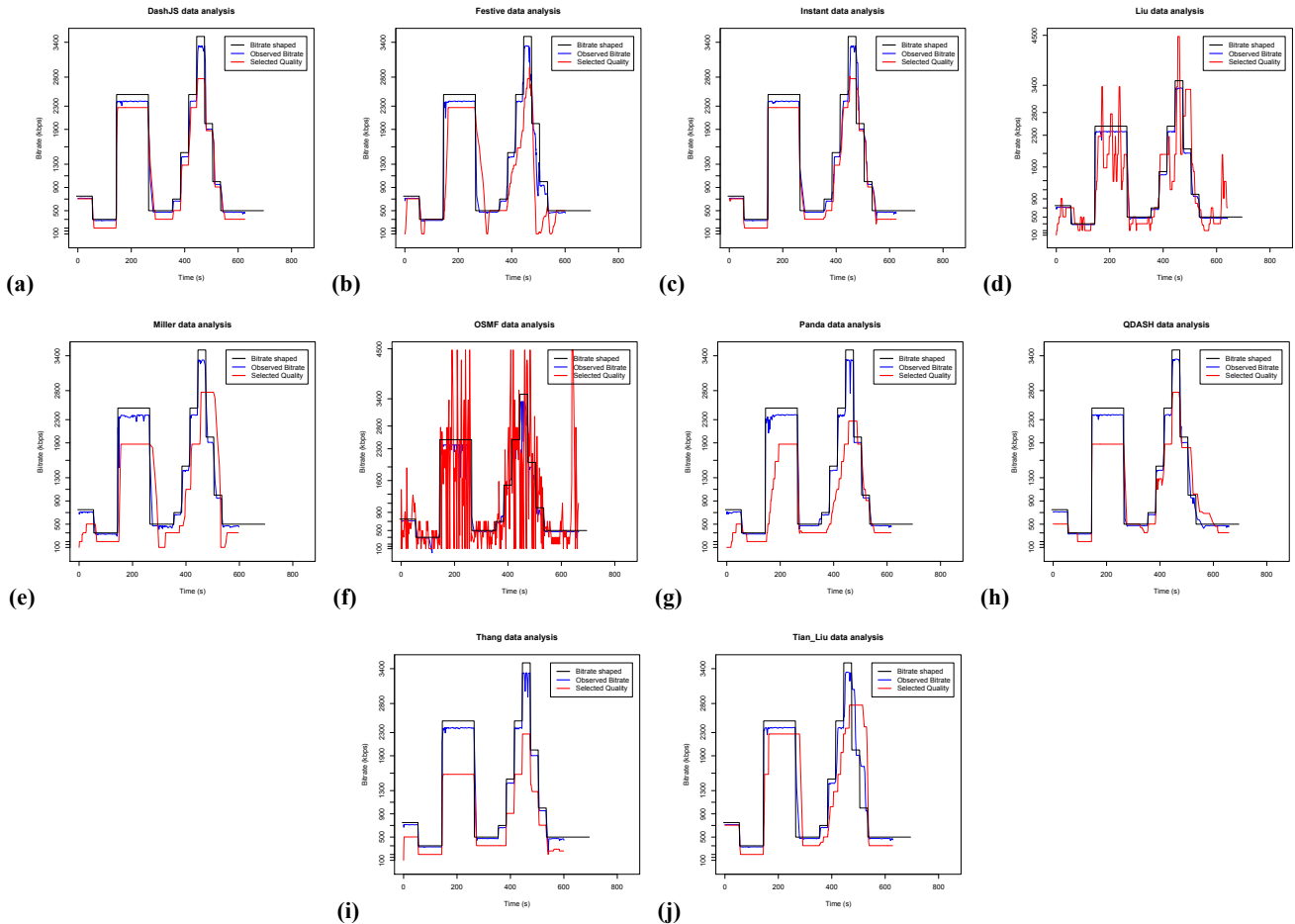

**Figure 4. Bandwidth Adaptation with 2s segment size for (a) DASH-JS, (b) FESTIVE, (c) Instant, (d) Liu, (e) Miller, (f) OSMF, (g) PANDA, (h) QDASH, (i) Thang, and (j) Tian-Liu.**

In addition to the QoE rating we gather various objective metrics such as number of stalls (i.e., buffer underruns), and the average media throughput of the client.

This methodology enables a subjective evaluation of different DASH adaptation logics within real-world environments as opposed to controlled environments and, thus, provides a more realistic evaluation of adaptive HTTP streaming systems. However, using crowdsourcing requires a more careful evaluation of the participant's feedback as outlined in [18]. Therefore, we filtered participants using browser fingerprinting, stimulus presentation time, actual QoE rating, and feedback from the pre-questionnaire as documented in [16].

In the following sections we provide the results of the objective and subjective evaluations.

## 4. OBJECTIVE RESULTS

We use the following metrics to compare the objective results of the adaptation logics in question. For the bandwidth adaptation we define the **bitrate shaped** as the bandwidth trajectory as shown in Figure 1, **observed bitrate** is the bitrate measured while downloading the segments (it provides the major basis and input for the adaptation logic), and **selected quality** corresponds to the representation selected as an output of the adaptation logic. Additionally, the **buffer level** provides the buffer fill state in seconds and the **autopause level** indicates buffer underruns/stalls.

A first comparison is the difference between segment lengths of 2s versus 10s and its impact on the buffer level. For brevity we show only results of one adaptation logic, i.e., DASH-JS, as conclusions on the segment length are similar for others. Figure 3 depicts the bandwidth adaptation and buffer level for DASH-JS using 2s (left side) and 10s (right side) segment length. The upper part of the figure clearly shows that shorter segment durations (2s) allow for better matching to the available bandwidth whereas longer durations (10s) enable smoother bandwidth adaptation, i.e., switches are not as abrupt as for shorter durations. However, the lower part reveals that longer segment size durations cause more stalls, specifically during sudden bandwidth changes with high amplitude (e.g., around second 350 and 600). Note that DASH-JS uses the MPD download for the initial bandwidth estimation and typically starts with higher quality representations that could lead to a higher start-up delay and eventually stalls (e.g., at the very beginning).

Figure 4 shows the bandwidth adaptation for all the adaptation logics with a segment size of 2s. Most of the adaptation logics follow the available bandwidth instantaneously and always select a representation lower than the observed bitrate except for Liu (d) and OSMF (f). The behavior of the former (Liu) can be explained due to its AIMD-like approach (Section 2.4), which always tries to increase the bitrate by additively selecting next higher quality representations until it exceeds the measured bandwidth (then followed by multiplicative decrease). The latter (OSMF) shows a somewhat unpredictable behavior but this has been reported independently already elsewhere [15][19] and is confirmed here also. Interestingly, the algorithms handle the start-up phase quite differently; some are conservative showing a step-wise behavior from the lowest representation (d, e, g) while others are more aggressive by switching to the appropriate representation right after the first few segments (b, i) or instantly selecting it right away (a, c, h, j).

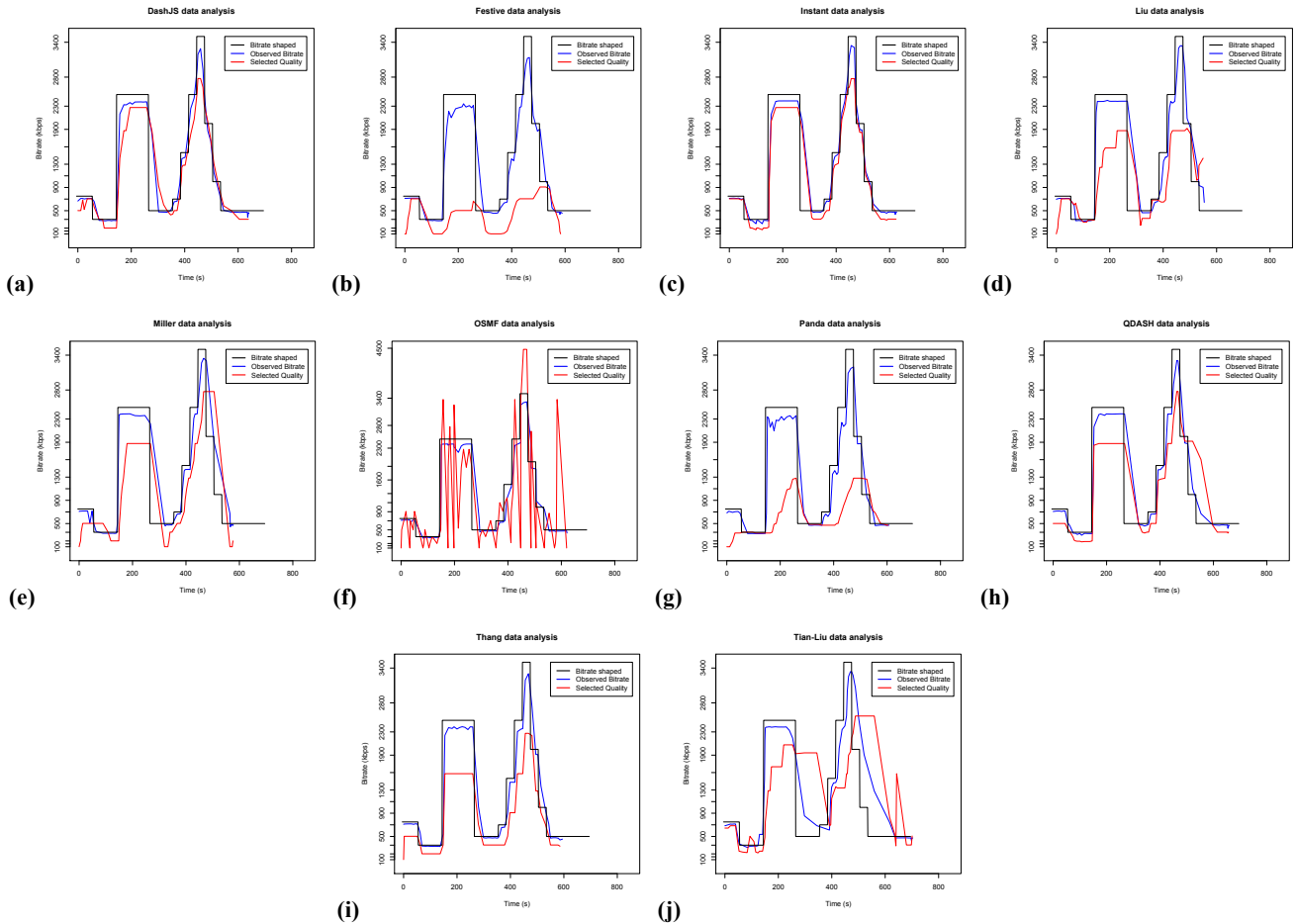

**Figure 5. Bandwidth Adaptation with 10s segment size for (a) DASH-JS, (b) FESTIVE, (c) Instant, (d) Liu, (e) Miller, (f) OSMF, (g) PANDA, (h) QDASH, (i) Thang, and (j) Tian-Liu.**

Figure 5 depicts the same results but with a segment size of 10s which reveals some interesting aspects if compared with Figure 4, except for OSMF which shows the same weird behavior although not to the same extent. In particular, FESTIVE (b) is much more conservative with 10s segment size compared to 2s. Representations higher than 700kbps are almost never selected resulting in a relatively low media throughput at the client. The algorithm Liu (d) performs now better than with 2s segment size and always stays below the observed bitrate. Miller (e), QDASH (h), and Thang (i) show roughly the same behavior and PANDA (g) is also more conservative but not as much as FESTIVE. Finally, Tian-Liu (j) seems to almost overcome the bandwidth drop in the middle of the trajectory but making a false estimation towards the end of streaming session.

Some of the behavior shown in Figure 4 and Figure 5 can be further analyzed by investigating the buffer level which is shown in Figure 6. For the 2s segment sizes (a-j), DASH-JS, FESTIVE, Instant, QDASH, and Tian-Liu provide a stable buffer and quite similar behavior. Interestingly, Miller, PANDA, and Thang have a much higher buffer fill level than others. The Liu algorithm as a very active, frequently changing buffer fill level while OSMF is very unpredictable. When looking at the results for the 10s segment sizes (k-t), DASH-JS, Instant, QDASH, and Tian-Liu are still unremarkable although they produce more stalls due to the larger segment size which becomes apparent during the bandwidth drops. The buffer of FESTIVE is now much higher due to the lower media throughput which allows for more data to be buffered. Liu is more stable than with 2s segment size which is also observed for OSMF but only to a certain extent (i.e., it is still unpredictable). Also Miller, PANDA, and Thang are comparable with the buffer when using the 2s segment sizes but with more stalls, specifically during bandwidth drops, which shows the impact of longer segment sizes.

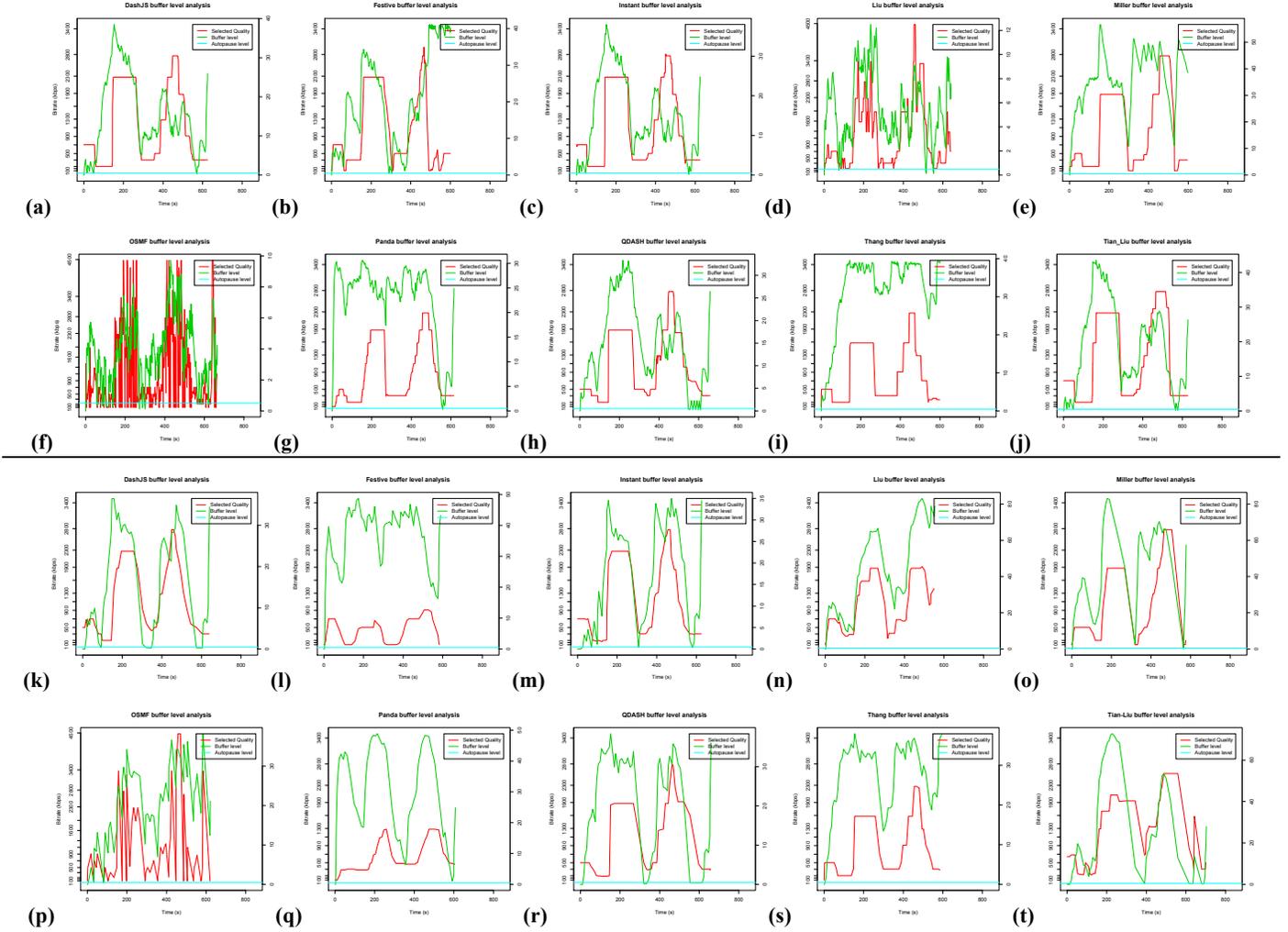

**Figure 6. Buffer Level with 2s and 10s segment size: (a-j) 2s segment size, (k-t) 10s segment size.**

Finally, we investigate the performance of the different adaptation logics using a set of predefined metrics as follows: inefficiency, instability, media throughput (mean of bitrates), buffer level, start time, and buffer underruns/stalls.

**Inefficiency** is defined according to Equation ( 2 ) and determines to what extent the algorithm utilizes the available network. The lower the value, the more efficiently the scheme is utilizing the network throughput in order to deliver the media content to the client device.

$$Inefficiency = \sum_{t} \frac{|b_{i,t} - W_{i,t}|}{W_{i,t}} \qquad (2)$$

where $b_{i,t}$ is the rate selected for the segment $i$ at time $t$ and $W_{i,t}$ is the observed bandwidth measured during downloading of the segments.

**Instability** provides the ratio between the observed switching steps and the sum of the selected bitrates over a window of k=20 seconds as shown in Equation ( 3 ).

$$Instability = \frac{\sum_{d=0}^{k-1} |b_{t-d} - b_{t-d-1}| \cdot \omega(d)}{\sum_{d=1}^{k} b_{t-d} \cdot \omega(d)} \qquad (3)$$

where the function $\omega(d) = k - d$ return a weight that adds more penalty to the most recent switches. A lower value for the instability means a smoother video quality adaptation to changing network conditions.

The **media throughput** is defined by the mean of bitrates that has been selected by the individual adaptation logics throughout the session, which is compared to the bitrate shaped (available bandwidth) and observed bitrate (measured while downloading the segments). The **buffer level** defines average buffer fill state in seconds throughout the streaming session and **start time** provides the time between the MPD request and until four seconds of media contents are available in the buffer. Finally, the number of **underruns/stalls** provides a very important metric for the user's Quality of Experience (QoE).

Figure 7 shows the performance results for the different segments sizes of 2s and 10s using media throughput, buffer level, underruns/stalls, and start-up delay.

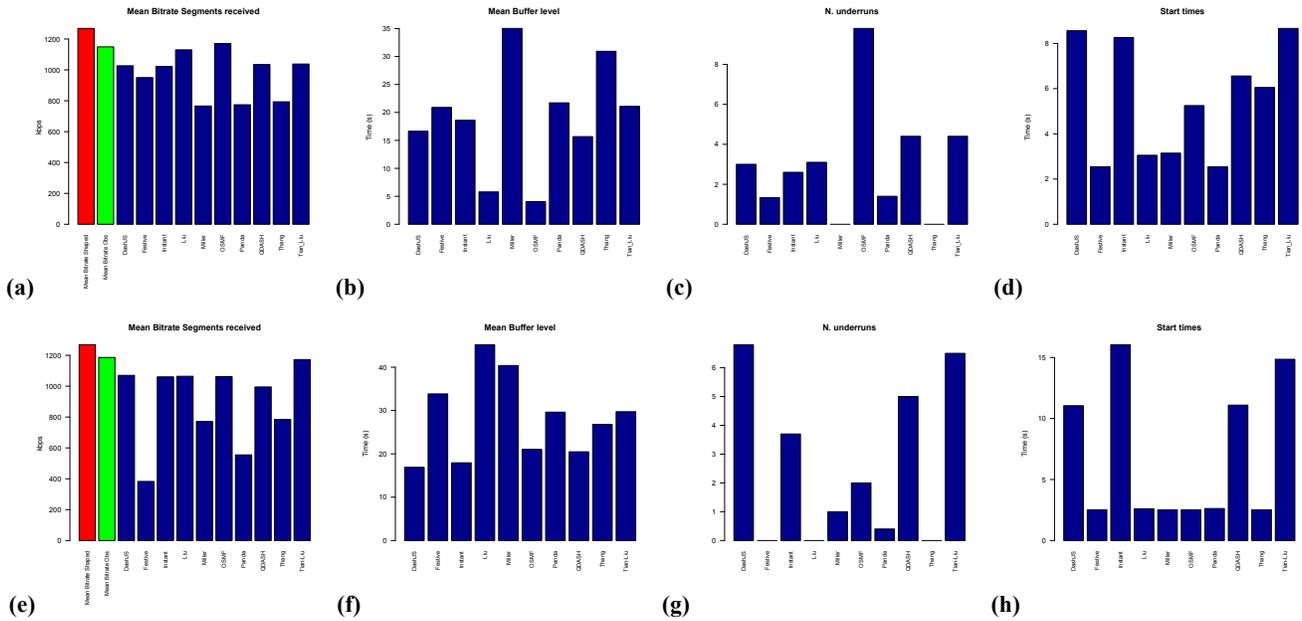

Figure 7. Performance Results for 2s (a-d) and 10s (e-h) Segment Size:
Media Throughput (a, e), Buffer Level (b, f), Buffer Underruns/Stalls (c, g), Start-up Delay (d, h).

The media throughput (a, e) is compared with the available bandwidth (red bar) and the measured bitrate (green bar). Interestingly, OSMF reaches the highest value (at least for 2s segment size) despite its unpredictable behavior but please note the results for the other metrics. Additionally, for 2s segment sizes, the algorithms Thang, Miller, and PANDA are below 800kbps on average whereas others are close to what has been observed while downloading the segments. Looking at the results of 10s segment size, FESTIVE (e) falls below 400kbps on average which is observed also in Figure 5(b).

A buffer level (b, f) greater than zero is maintained by all the adaptation logics in all the cases and it is larger than 15s of buffered segments for most of the cases. Only Liu and OSMF have a relatively low buffer fill state for the 2s segment size which results in a high number of stalls for OSMF (c). For Liu this is also reflected in the adaptation behavior as shown in Figure 4(d) where segment bitrates often exceed the available bandwidth. However, Liu and also FESTIVE have a much higher buffer fill state using a segment size of 10s which adopts a more conservative adaptation behavior than with 2s. PANDA is also more conservative in the 10s case but this does not impact the buffer fill state. Others show roughly the same buffer level for both cases.

The buffer underruns/stalls (c, g) represent a very important QoE metric and only Thang manages to avoid stalls in both cases. Interestingly, some approaches manage to reduce the number of stalls when using 10s segments which could be explained due to an increased buffer fill state (e.g., FESTIVE and Liu). However, others like DASH-JS, Instant, QDASH, and Tian-Liu result in an increased number of stalls when using 10s segments which is due to the already low buffer fill state.

Finally, the start-up delay (d, h) is low in general but expectably higher for those approaches which use the MPD download for estimating the bitrate of the first segment to be retrieved like DASH-JS or Instant. Thus, using the MPD download as an educated guess for selecting the initial representation maybe only used for use cases where start-up delay does not play an important role, e.g., for on-demand content of full length videos (like Netflix), as opposed to short video clips (e.g., YouTube) where a high start-up delay is usually not tolerable.

A summary of performance results for the media throughput and buffer underruns/stalls is given in Table 2.

The results for the derived metrics – inefficiency and instability – are depicted in Figure 8. Notably, the inefficiency metric increases for FESTIVE in case 10s segments are used which is reflected also in the low media throughput. Nevertheless, the instability is still very low as it can be also seen in the low number of stalls. The results for OSMF are also reflected in the performance metrics above and others show similar results for both 2s and 10s segment sizes. In particular, DASH-JS and Instant provide a consistently low inefficiency and instability among all tested adaptation logics.

Table 2. Summary of Results: Media Throughput and Stalls for 2s and 10s segment sizes.

|  | Throughput 2s | Throughput 10s | Stalls 2s | Stalls 10s |
|---|---|---|---|---|
| *Avail. Bw.* | 1,269.53 | | – | – |
| *Measured Bw.* | 1,194.07 | 1,252.87 | – | – |
| **DASH-JS** | 1,026.52 | 1,069.95 | 3 | **6.8** |
| **FESTIVE** | 950,10 | **382.69** | 1.33 | **0** |
| **Instant** | 1,022.54 | 1,060.11 | 2.6 | 3.7 |
| **Liu** | 1,129.69 | 1,063.92 | 3.1 | **0** |
| **Miller** | 766.27 | 770.91 | **0** | 1 |
| **OSMF** | **1,170.65** | 1,061.79 | **9.8** | 2 |
| **PANDA** | **774.18** | 554.89 | 1.4 | 0.4 |
| **QDASH** | 1,034.71 | 994.67 | 4.4 | 5 |
| **Thang** | 793.29 | 783.93 | **0** | **0** |
| **Tian-Liu** | 1,037.71 | **1,172.28** | 4.4 | 6.5 |

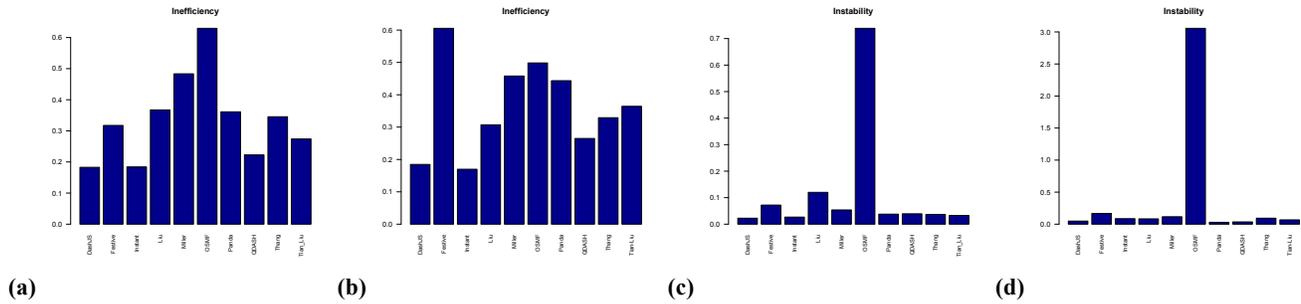

**Figure 8. Inefficiency and Instability for (a-b) 2s segment size and (c-d) 10s segment size.**

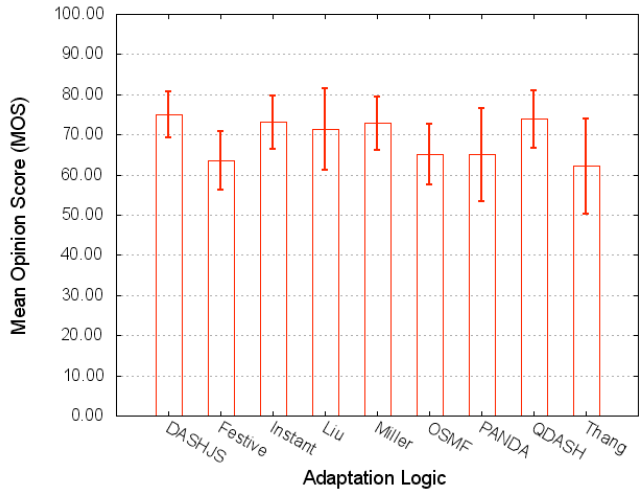

**Figure 9. Mean Opinion Score (MOS) per Adaptation Logic with a 95% Confidence Interval.**

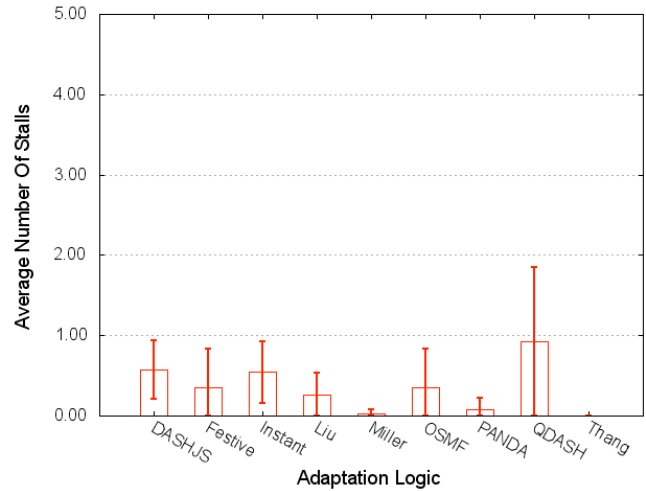

**Figure 10. Average Number of Buffer Underruns/Stalls per Adaptation Logic with a 95% Confidence Interval.**

## 5. SUBJECTIVE RESULTS

In total 220 microworkers participated in the subjective quality assessment from which 19 participants were excluded from the evaluation (due to issues during the crowdsourcing test as outlined in Section 3.3). From the remaining 201 participants were 143 male and 58 female with an average age of 28.

The results presented in this section reflect the behavior of the adaptation logics in a real-world environment with subjects spread across Europe accessing the test sequences over the open Internet.

Unfortunately, the crowdsourcing study did not provide any data for the algorithm from Tian-Liu due to a software error and, thus, this algorithm is excluded from the subjective results.

Figure 9 depicts the QoE in terms Mean Opinion Score (MOS) per adaptation logic (95% confidence interval). Interestingly, DASH-JS (and also Instant) provides the highest MOS value but due to overlapping confidence intervals relatively little can be stated whether it performs significantly better than the other algorithms. However, it provides a good indication about its effectiveness in a real-world environment. OSMF does not have the lowest MOS value despite its worse performance during the objective evaluation. In particular, Thang has the lowest MOS value although – during the objective evaluation – it does not cause any stalls but comes with a relatively low media throughput for both segment sizes.

In addition to the QoE results, we also collect objective performance metrics that predominantly impact the QoE. In particular, we present results for the number of buffer underruns/stalls and media throughput as observed in such a real-world environment, which are depicted in Figure 10 and Figure 11, respectively. In general, the number of stalls is (very) low on average with some outliers for QDASH (i.e., "larger" confidence interval than others but still relatively small). Thus, all adaptation logics tested within a real-world environment basically confirm to the most important QoE guideline for adaptive HTTP streaming, i.e., a zero/low number of stalls. Regarding the average media throughput we observe lower values for FESTIVE, Miller, PANDA, and Thang which confirm results obtained during the objective evaluation within a controlled environment (cf. Table 2). On the other hand, DASH-JS provides the highest media throughput, which is again an indication that simplicity rules out complexity in terms of performance.

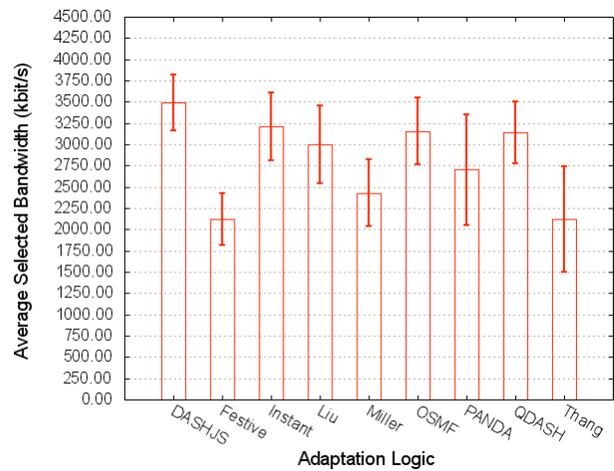

**Figure 11. Average Media Throughput per Adaptation Logic with a 95% Confidence Interval.**

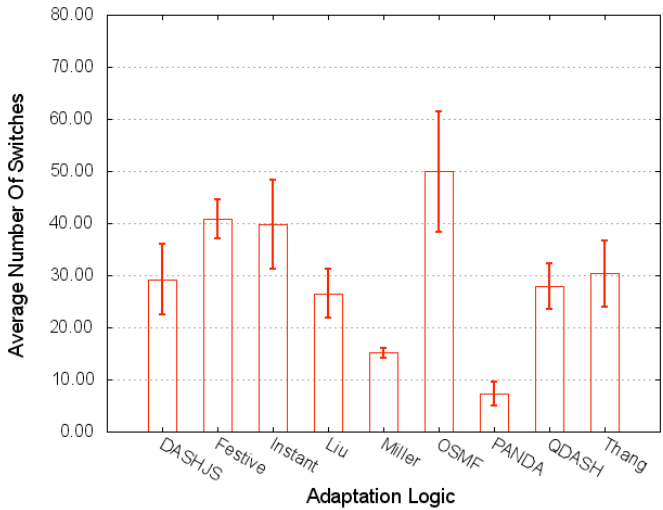

**Figure 12. Average Number of Representation Switches per Adaptation Logic with a 95% Confidence Interval.**

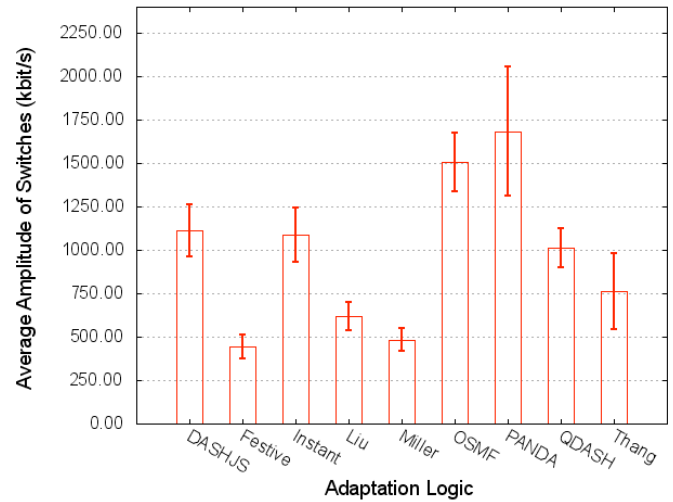

**Figure 13. Average Amplitude of Representation Switches per Adaptation Logic with a 95% Confidence Interval.**

We also captured the number of representation switches per adaptation logic which are shown in Figure 12. Additionally, Figure 13 shows the average amplitude of the switches (i.e., the distance between the representation switches). OSMF has the highest number of switches and second highest amplitude of switches which is also apparent when looking at Figure 4 and Figure 5. FESTIVE has the second highest number of switches – slightly "ahead" of Instant – but at a very low amplitude which indicates switches only among neighboring representations at a lower rate (cf. low media throughput of FESTIVE in Figure 11).

Miller and PANDA have the lowest number of quality switches; the former has also the second lowest switching amplitude whereas the latter has the highest amplitude. Thus, one may conclude that PANDA does not switch very often but, when switching is required, it switches to the right representation without much fine-tuning. The adaptation logic with the highest MOS and media throughput – DASH-JS – provides an average performance in terms of number of switches and amplitude compared to all others. Interestingly, its number of switches is lower than Instant while performing almost equal regarding the switching amplitude thanks to its weighting factors.

However, in general, the switching amplitude needs to be considered always together with the number of switches in order to draw any conclusions (cf. PANDA in Figure 12 and Figure 13).

## 6. CONCLUSIONS

In this paper we have investigated various adaptation HTTP streaming adaptation logics/algorithms proposed in the literature. We provide a comprehensive evaluation using both objective and subjective metrics within both controlled and real-world environments. The subjective results gathered in real-world environments using crowdsourcing confirm the objective results conducted within a controlled environment. As somehow expected, there is not clear winner which takes it all but, interestingly, simple approaches – like DASH-JS and Instant – perform reasonably well in both cases. Therefore, we can conclude that these simple solutions clearly follow Einstein's rule to make things as simple as possible but not simpler.

The methodology adopted in this paper can be easily reused for both objective evaluations in controlled environments and subjective evaluations in real-world environments. In particular, it allows for an easy, fast, and reliable evaluation of adaptive HTTP streaming systems.

Future work in this area comprises further evaluating the data gathered during these evaluations and providing means to test different network delays (round trip time) and competing clients adopting different approaches (i.e., different adaptation logics and other traffic). Furthermore, performing such experiments in different contexts (e.g., home vs. mobile) would reveal additional results regarding the usability of the different adaptation logics. Therefore, a more automated setup for conducting and the analysis of such experiments with pluggable adaptation logics is hereby solicited.

## 7. ACKNOWLEDGMENTS

This work was supported in part by the EC in the context of the SocialSensor (FP7-ICT-287975) and QUALINET (COST IC 1003) projects and partly performed in the Lakeside Labs research cluster at AAU.